\documentclass[a4, conference]{IEEEtran}
\usepackage{setspace}
\usepackage{cite}
\usepackage{pgfplots}

\pgfplotsset{compat=1.10}

\usepgfplotslibrary{fillbetween}

\usepackage{bm}
\usepackage{lipsum}
\usepackage{makeidx}
\usepackage{enumerate}
\usepackage{color}
\usepackage{cite}
\usepackage{amsmath,amsthm}   
\usepackage{amssymb}
\usepackage{nomencl}
\usepackage{multirow}
\usepackage{graphicx}
\usepackage{epstopdf}
\usepackage{threeparttable}
\usepackage{multicol}
\DeclareGraphicsExtensions{.pdf,.jpeg,.png,.jpg,.emf,.eps}
\usepackage{calc}
\usepackage{here}
\usepackage{url}

\usepackage{upref}
\usepackage{comment}
\usepackage{times}
\usepackage{dsfont}
\usepackage{epic,eepic}
\usepackage{rawfonts}
\usepackage[T1]{fontenc}
\usepackage{latexsym}
\usepackage{amsfonts}

\usepackage{capitalgreekitalic}

\usepackage[font=small,labelfont=bf]{caption}
\usepackage[font=small,labelfont=bf]{subcaption}

\usepackage{xcolor}
\usepackage{tikz,pgfplots}
\usetikzlibrary{calc}
\usetikzlibrary{intersections}
\usetikzlibrary{arrows,shapes}
\usetikzlibrary{positioning}

\theoremstyle{definition}

\IEEEspecialpapernotice{(Invited Paper)}

\allowdisplaybreaks

\begin{document}

\title{Energy Efficiency Comparison of RIS Architectures in MISO Broadcast Channels
}
\author{
   \IEEEauthorblockN{Mohammad Soleymani$^1$, Ignacio Santamaria$^2$, Eduard Jorswieck$^3$, Marco Di Renzo$^4$, Jesús Gutiérrez$^5$}
   \IEEEauthorblockA{$^1$Signal \& System Theory Group, Universit\"at Paderborn, Germany \\
   $^2$Dept. Communications Engineering, Universidad de Cantabria, Spain\\
   $^3$ Institute for Communications Technology, Technische Universit\"at Braunschweig, Germany\\
   $^4$Université Paris-Saclay, CNRS, CentraleSupélec, Laboratoire des Signaux et Systèmes, 91192 Gif-sur-Yvette, France\\
   $^5$IHP - Leibniz-Institut für Innovative Mikroelektronik, 15236 Frankfurt (Oder), Germany\\
                     Email: \small{\protect\url{mohammad.soleymani@uni-paderborn.de}}, \small{\protect\url{i.santamaria@unican.es}}, \small{\protect\url{jorswieck@ifn.ing.tu-bs.de}}\\
                     \small{\protect\url{marco.di-renzo@universite-paris-saclay.fr}}, \small{\protect\url{teran@ihp-microelectronics.com}},
}
}

\maketitle
\begin{abstract}
In this paper, we develop energy-efficient schemes for multi-user multiple-input single-output (MISO) broadcast channels (BCs), assisted by reconfigurable intelligent surfaces (RISs). To this end, we consider three architectures of RIS: locally passive diagonal (LP-D), globally passive diagonal (GP-D), and globally passive beyond diagonal (GP-BD). In a globally passive RIS, the power of the output signal of the RIS is not greater than its input power, but some RIS elements can amplify the signal. In a locally passive RIS, every element cannot amplify the incident signal. We show that these RIS architectures can substantially improve energy efficiency (EE) if the static power of the RIS elements is not too high. Moreover, GP-BD RIS, which has a higher complexity and static power than LP-D RIS and GP-D RIS, provides better spectral efficiency, but its EE performance highly depends on the static power consumption and may be worse than its diagonal counterparts.
\end{abstract} 

\section{Introduction}\label{1}
Energy efficiency (EE) has been among the main concerns of wireless communication systems for more than a decade \cite{buzzi2016survey}. The high energy expenses and the global warming make energy-efficient systems even more relevant in the future, where one of the primary goals of 6G networks is to be around 100 times more energy-efficient than 5G systems \cite{wang2023road, gong2022holographic}. To achieve this goal, 6G has to utilize promising technologies like reconfigurable intelligent surfaces (RISs), which can control the signal propagation, leading to coverage enhancement, interference reduction, in a nearly passive manner, hence enhancing the EE \cite{di2020smart, wu2021intelligent}.

The EE performance of RISs has been studied in \cite{huang2019reconfigurable, wu2019intelligent, soleymani2022improper, soleymani2024optimization, xu2022robust, ihsan2022energy, soleymani2022noma, soleymani2023noma, soleymani2022rate, soleymani2023energy, huang2022integrated, soleymani2023optimization},
where it was shown that RISs can substantially enhance the EE under different network assumptions. For instance, the authors of \cite{huang2019reconfigurable} showed that an RIS enhances
the global EE (GEE) of multiple-input single-output (MISO)
broadcast channels (BCs). The authors of \cite{wu2019intelligent} showed that RISs reduce the power consumption, while ensuring minimum rate per user. In \cite{soleymani2022improper, soleymani2024optimization}, it was shown that RISs increase the minimum EE of the users and the GEE of the system in multi-cell multiple-input multiple-output (MIMO) BCs. The authors of \cite{xu2022robust} proposed robust designs to improve the EE of heterogeneous networks aided by RISs. The authors of \cite{huang2022integrated} showed that an RIS can enhance the EE of an integrated sensing and communication system.

In \cite{huang2019reconfigurable, wu2019intelligent, soleymani2022improper, xu2022robust, ihsan2022energy, huang2022integrated}, an RIS with a diagonal scattering matrix and a locally passive (LP) constraint was considered. However, there exist other RIS architectures with different power supply assumptions and circuit designs. The performance of an RIS can be enhanced in some scenarios if the RIS elements are connected to a power amplifier, which allows the
RIS to increase the scattered power. This solution is known
as the active RIS \cite{zhang2022active}. Although an active RIS is a promising technology, it needs a power supply, which may not be possible
in some practical scenarios. In contrast, nearly passive RISs
do not require a separate power supply for power amplification
and can operate almost autonomously. Hence, in this work, we
focus on different architectures for nearly passive RISs.

When an RIS operates in a nearly passive mode, its total output power is less than or equal to the total input power. There can be two possibilities to realize a nearly passive RIS. The first possibility is to operate each RIS element in a nearly passive mode, referred to as the LP RIS, where the amplitude of the reflection coefficient of each RIS element is no greater than one. The other possibility is to allow some elements
to amplify and attenuate the received signal as long as the
total power of the reflected signal is not greater than the total
power of the incident signal \cite{fotock2023energy, fotock2023energy2, fotock2023energy3}. This RIS architecture is referred to as globally passive (GP) RIS. It was shown in \cite{fotock2023energy} that a GP RIS can substantially outperform a LP RIS in terms of GEE in a SIMO multiple-access channel (MAC).

To improve the performance of a nearly passive RIS (either LP or GP), its elements can be connected to realize a non-diagonal scattering matrix, referred to as BD-RIS architecture \cite{li2023reconfigurable, li2022beyond}. Making the matrix of the RIS coefficients non-diagonal
enlarges the set of optimization parameters, which provides additional degrees of freedom at the cost of higher computational and implementation complexities. In \cite{nerini2023closed, santamaria2023snr}, it was shown that the BD-RIS architecture increases the signal-to-noise ratio (SNR) of a point-to-point single- and multiple-antenna system. Note that BD-RIS is more general than a diagonal RIS (D-RIS), and an optimized BD-RIS architecture never performs worse than any D-RIS design from a spectral efficiency point of view. However, a BD-RIS architecture has a more sophisticated circuit design with a significantly larger number of circuit elements, which consumes much more static power than a D-RIS architecture. Thus, a BD-RIS may not be necessarily more energy-efficient than a D-RIS architecture due to the higher static power required to operate it.

In this paper, we propose energy-efficient designs for the three considered architectures of reflective RIS, i.e., LP-D, GP-D, and GP-BD. In \cite{fotock2023energy}, it was shown that active and GP D-RISs can increase the global EE of an uplink (UL) multi-user SIMO system. In this paper, we consider the GP-BD RIS architecture in addition to the LP-D RIS and GP-D RIS architectures, and maximize the minimum EE of a downlink (DL) multi-user MISO BC. We show that the three considered RIS architectures can substantially improve the max-min EE of the MISO BC if the static power consumption of the RIS architectures is not high, which is a realistic assumption for a nearly passive (either locally or globally) architecture. The GP-BD architecture outperforms the LP-D and GP-D architectures when the static power consumption is not too high. However, as the static power of each RIS/circuit element grows, the GP-BD RIS consumes much more power than the LP-D and GP-D architectures, and thus the EE of the GP-BD architecture may become less than the EE of the LP-D and GP-D architectures.

\section{System Model}
We consider a BC with a $K$-antenna base station (BS) serving $L$ single-antenna users. We assume that an $N$-element RIS assists the BS to serve the users. We consider three RIS architectures, i.e., LP-D, GP-D and GP-BD. In all of these architecture, the channel
between the BS and user $l$ is
\begin{equation}\label{(1)}
    {\bf h}_l ({\bf \Psi}) ={\bf f}_l {\bf \Psi} {\bf F}+{\bf g}_l \in\mathbb{C}^{1\times K},
\end{equation}
where ${\bf F}$ is the channel between the BS and the RIS, ${\bf f}_l$ is the channel between the RIS and user $l$, ${\bf g}_l$ is the channel between the BS and user $l$, and ${\bf \Psi}$ is the matrix of the RIS reflection coefficients. In an LP-D RIS, ${\bf \Psi}$ is a diagonal matrix in which the absolute value of each diagonal element is equal to one. Therefore, the set of feasible values of ${\bf \Psi}$ is
\begin{equation}\label{(2)}
    \mathcal{D}_{LPD}=\{\psi_{nj}: |\psi_{nn}|=1, \psi_{nj}=0, \forall n \neq j \}  ,
\end{equation}
where $\psi_{nj}$ is the entry in the $n$-th row and $j$-th column of ${\bf \Psi}$.

When we employ a GP RIS architecture (either diagonal or not), the power of the reflected signal needs to be less than or equal to the power of the incident signal. In analytical terms, we obtain
\begin{equation}\label{(3)}
    p_{out}-p_{in}=\text{Tr}(\mathbb{E}\{{\bf t}{\bf t}^H\} ({\bf \Psi}^H {\bf \Psi} -{\bf I}_N ) )\leq 0,
\end{equation}
where ${\bf t}$ is the incident signal, with power $\text{Tr}(\mathbb{E}\{{\bf t}{\bf t}^H\})$, $\text{Tr}({\bf \Psi}\mathbb{E}\{{\bf t}{\bf t}^H\}{\bf \Psi}^H)$ is the power of the reflected signal, $\mathbb{E}\{{\bf t}\}$ denotes the mathematical expectation of ${\bf t}$, $\text{Tr}({\bf X})$ denotes the trace of the matrix ${\bf X}$, and ${\bf I}_N$ denotes an $N\times N$ identity matrix. Thus, the feasibility set of ${\bf \Psi}$ for the GP-D RIS architecture is
\begin{equation}\label{(4)}
    \mathcal{D}_{GPD}\!\!=\!\{\!{\bf \Psi}\!\!:\!
    \text{Tr} (\mathbb{E}\{\!{\bf t}{\bf t}^H\}\! (\!{\bf \Psi}^H {\bf \Psi} \!-\!{\bf I}_N ) )\!\leq\! 0, \psi_{nj}\!=\!0, \forall n \!\neq\! j \}.\!
\end{equation}
Moreover, the feasibility set of ${\bf \Psi}$ for GP-BD RIS is
\begin{equation}\label{(5)}
    \mathcal{D}_{GPBD}\!=\!\{{\bf \Psi}\!:{\bf \Psi}={\bf \Psi}^T\!, \text{Tr} (\mathbb{E}\{{\bf t}{\bf t}^H\} ({\bf \Psi}^H {\bf \Psi} \!-{\bf I}_N ) )\!\leq\! 0 \}.
\end{equation}
Note that we consider a symmetric matrix for GP-BD RIS, since implementing beyond diagonal RISs with a symmetric design is much simpler \cite{li2022beyond}. However, one can easily extend the model and the analytical scheme to a GP-BD RIS with a non-symmetrical ${\bf \Psi}$. The sets $\mathcal{D}_{GPD}$ and $\mathcal{D}_{GPBD}$ are convex in ${\bf \Psi}$; however, $\mathcal{D}_{LPD}$ is non-convex. Moreover,
$\mathcal{D}_{LPD}\subset\mathcal{D}_{GPD}\subset\mathcal{D}_{GPBD}$, which means that GP-BD RIS is more general than the two other RIS architectures, and an optimal GP-BD RIS scheme does not perform worse than any LP and GP diagonal RIS from a spectral efficiency perspective.
Hereafter, we denote the feasibility set of ${\bf \Psi}$ as $\mathcal{D}$ unless we explicitly refer to one of the sets $\mathcal{D}_{LPD}$, $\mathcal{D}_{GPD}$ or $\mathcal{D}_{GPBD}$.
\hspace{-.8cm}

\subsection{Signal model and problem formulation}
\hspace{-.15cm}
The transmit signal of the BS is
\begin{equation}\label{(6)}
   {\bf x}=\sum_{l=1}^L{\bf w}_ls_l\in \mathbb{C}^{K\times 1},
\end{equation}
where ${s}_l\sim \mathcal{CN}(0,1)$ is the data message intended for user $l$, which is independent of  ${s}_j$ for $j\neq l$, and ${\bf w}_l$ is the beamforming vector corresponding to  ${s}_l$. Thus, the signal ${\bf x}$ has zero mean and covariance matrix ${\bf X} = \mathbb{E}\{{\bf x}{\bf x}^H\}=\sum_l {\bf w}_l{\bf w}_l^H$. Additionally, the transmit power of the BS is $\text{Tr}({\bf X}) =\sum_l {\bf w}_l^H{\bf w}_l\leq P$, where $P$ is the power budget of the BS.

The received signal at the RIS is ${\bf t} = {\bf F}{\bf x}$, and $\mathbb{E}\{{\bf t}{\bf t}^H\} =
{\bf F}(\sum_l {\bf w}_l{\bf w}_l^H){\bf F}^H$. Additionally, the output signal of the RIS
is 	${\bf \Psi} {\bf F}{\bf x}$, where ${\bf F}$ is defined in \eqref{(1)}. Thus, the GP constraint in
\eqref{(3)} can be written as
\begin{equation}\label{(7)}
    \text{Tr}\left(
    \sum_l {\bf F} {\bf w}_l{\bf w}_l^H {\bf F}^H
    ({\bf \Psi}^H{\bf \Psi}-{\bf I}_N)
    \right)
    \leq 0.
\end{equation}
Moreover, the received signal at user $l$ is
\begin{equation}\label{(8)}
    { y}_l={\bf h}_l({\bf \Psi}) \sum_{j=1}^L{\bf w}_js_j+n_l,
\end{equation}
where ${n}_l\sim \mathcal{CN}(0,\sigma^2)$ is the additive white Gaussian noise. We assume that each user treats interference as noise to decode its own signal. Thus, the rate of user $l$ is
\begin{equation}\label{(9)}
    r_l=\log\left(
    1+\frac{|{\bf h}_l({\bf \Psi} ){\bf w}_l |^2}{\sigma^2 +\sum_{j\neq l} |{\bf h}_l({\bf \Psi} ){\bf w}_j |^2}
    \right).
\end{equation}
Additionally, the EE of user $l$ is
\begin{equation}\label{(10)}
    e_l=\frac{r_l}{P_c+\beta{\bf w}_l^H{\bf w}_l}
    =\frac{\log\left(
    1+\frac{|{\bf h}_l({\bf \Psi} ){\bf w}_l |^2}{\sigma^2 +\sum_{j\neq l} |{\bf h}_l({\bf \Psi} ){\bf w}_j |^2}
    \right)}{P_c+\beta{\bf w}_l^H{\bf w}_l},
\end{equation}
where $\beta^{-1} $ is the power efficiency of the BS, ${\bf w}_l^H{\bf w}_l$ is the power of the transmitted signal for user $l$, and $P_c$ is the static
operational power per user, given by
\begin{equation}\label{(11)}
   P_c=\frac{P_{BS}+P_{RIS}}{L}+P_{UE}=P_t+\frac{P_{RIS}}{L},
\end{equation}
where $P_{BS}$, $P_{RIS}$, and $P_{UE}$ are the static power consumption
to operate the BS, the RIS and each user equipment, respectively.
The static power of the RIS depends on the considered architecture and the number of circuit/RIS elements. For a nearly passive diagonal RIS architecture, we have
\begin{equation}\label{(12)}
    P_{RIS}=P^{D}_{RIS,0}+NP^{D}_{RIS,n},
\end{equation}
where $P^{D}_{RIS,0}$ is the static power of the diagonal architecture,
and $P^{D}_{RIS,n}$ is the static power of each RIS element, and $N$
is the number of RIS elements, which coincides with the number of circuit elements in this case. Moreover, the static power
of a GP-BD RIS is
\begin{equation}\label{(13)}
    P_{RIS}=P^{BD}_{RIS,0}+N_cP^{BD}_{RIS,n},
\end{equation}
where $P^{BD}_{RIS,0}$ is the static power of the BD architecture, and $P^{BD}_{RIS,n}$ is the static power of each circuit element for implementing a BD-RIS, and $N_c$ is the number of circuit elements of GP-BD RIS. In this paper, we consider a fully-connected architecture for GP-BD RIS, which has $N_c = N(N-1)/2$ circuit elements. Note that the rate and EE of each user are functions of $\{{\bf w}\}=\{{\bf w}_l:\forall l\} $, and 	${\bf \Psi}$.
However, we omit the dependency to simplify the writing.

In this paper, we aim at maximizing the minimum EE of the users by solving the following problem
\begin{subequations}\label{(14)}
\begin{align}    
   \max_{ \{{\bf w}\}, {\bf \Psi}\in \mathcal{D} }\, & \min_l \{e_l\}& \text{s.t.}\,\,&r_l\geq r_l^{th},\,\,\forall l,
   \\
   &&&\sum_l {\bf w}_l^H{\bf w}_l\leq P,
    \end{align}
\end{subequations}
where $r_l^{th}$ is the minimum required rate of user $l$. The optimization problem in \eqref{(14)} is non-convex and falls into the category of fractional
programming (FP) problems. We refer to the solution
of \eqref{(14)} as the max-min EE.

\section{Proposed energy-efficiency schemes}
To derive a suboptimal solution for \eqref{(14)}, we utilize iterative numerical optimization tools like alternating optimization (AO) and the generalized Dinkelbach algorithm (GDA). We start with a feasible initial point $\{{\bf w}^{(0)}\} $, ${\bf \Psi}^{(0)}$ and then update
either $\{{\bf w}\}$ or ${\bf \Psi}$	 at each step. Specifically, at the iteration $m$, we first update  $\{{\bf w}\}$ by solving \eqref{(14)} while ${\bf \Psi}$	 is kept fixed to ${\bf \Psi}^{(m-1)}$. Then to update 	${\bf \Psi}$, we solve \eqref{(14)} when  $\{{\bf w}\}$ is kept fixed to $\{{\bf w}^{(m)}\}$. We iterate this procedure until convergence.

\subsection{Optimizing $\{{\bf w}\}$}
When ${\bf \Psi}$ is kept fixed to ${\bf \Psi}^{(m-1)}$, \eqref{(14)} can be written as
\begin{subequations}\label{(15)}
\begin{align}    
   \max_{ \{{\bf w}\}}\, & \min_l \{e_l\}& \text{s.t.}\,\,&r_l\geq r_l^{th},\,\,\forall l,
   \\
   &&&\sum_l {\bf w}_l^H{\bf w}_l\leq P,
    \end{align}
\end{subequations}
which is an FP problem. To calculate a suboptimal solution
for \eqref{(15)}, we first obtain appropriate lower bounds for $r_l$, using
the bound in \cite[Eq. (71)]{soleymani2023spectral}, which yields
\begin{multline}\label{(16)}
r_l\geq \tilde{r}_l (\{{\bf w}\})
={r}_l (\{{\bf w}^{(m-1)}\},{\bf \Psi}^{(m-1)})-a_l
\\
+\frac{2\mathfrak{R}\left\{
\left( {\bf h}_l({\bf \Psi}^{(m-1)} ) {\bf w}_l^{(m-1)}
\right)^*
{\bf h}_l({\bf \Psi}^{(m-1)} ){\bf w}_l
\right\}
}{\sigma^2 +\sum_{j\neq l} |{\bf h}_l({\bf \Psi}^{(m-1)} ){\bf w}_j^{(m-1)} |^2}
\\
-a_l\frac{\sigma^2 +\sum_{j\neq l} |{\bf h}_l({\bf \Psi}^{(m-1)} ){\bf w}_j |^2}{\sigma^2 +\sum_{j\neq l} |{\bf h}_l({\bf \Psi}^{(m-1)} ){\bf w}_j^{(m-1)} |^2},
\end{multline}
where $\mathfrak{R}(x)$ denotes the real part of the complex
variable $x$, and
\begin{equation}
  a_l=\frac{|{\bf h}_l({\bf \Psi}^{(m-1)} ){\bf w}_l^{(m-1)} |^2}{\sigma^2 +\sum_{j\neq l} |{\bf h}_l({\bf \Psi}^{(m-1)} ){\bf w}_j^{(m-1)} |^2}.  
\end{equation} 
Replacing $r_l$ with $\tilde{r}_l$ for all $l$ gives the non-convex optimization problem
\begin{subequations}\label{(17)}
\begin{align}    
   \max_{ \{{\bf w}\}}\, & \min_l \{\tilde{e}_l\}& \text{s.t.}\,\,& \tilde{r}_l\geq r_l^{th},\,\,\forall l,
   \\
   &&&\sum_l {\bf w}_l^H{\bf w}_l\leq P,
    \end{align}
\end{subequations}
where $\tilde{e}_l=\frac{\tilde{r}_l}{P_c+\beta{\bf w}_l^H{\bf w}_l}$. Since $\tilde{r}_l$ and the denominator of $\tilde{e}_l$ are,
respectively, concave and convex in $\{{\bf w}\}$, we can derive the
optimum of \eqref{(17)} by using the GDA. Specifically, the global optimal
solution of \eqref{(17)} can be iteratively calculated by solving
\begin{subequations}\label{(18)}
\begin{align}    
   \max_{ \{{\bf w}\}}\, & \min_l \{\tilde{r}_l - \eta^{(t)} (P_c+\beta{\bf w}_l^{H}{\bf w}_l) \} \\
   \text{s.t.}\,\,& \tilde{r}_l\geq r_l^{th},\,\,\forall l,
   \\
   &\sum_l {\bf w}_l^H{\bf w}_l\leq P,
    \end{align}
\end{subequations}
and updating $\eta^{(t)}$ as
\begin{equation}\label{(19)}
    \eta^{(t)}=\min_l \left\{ 
    \frac{r_l\left(\{{\bf w}^{(t-1)}\} \right)}{P_c+\beta{\bf w}_l^{(t-1)
    ^H}{\bf w}_l^{(t-1)}}
    \right\},
\end{equation}
where $t$ is the index of the GDA iteration, and $\{{\bf w}^{(t-1)}\}$ is
the solution of \eqref{(18)} at the iteration $t-1$.

\subsection{Optimizing ${\bf \Psi}$}
When $\{{\bf w}\}$ is kept fixed to $\{{\bf w}^{(m-1)}\}$, \eqref{(14)} can be written as
\begin{align}\label{(20)}    
   \max_{ {\bf \Psi}\in\mathcal{D}}\, & \min_l \{e_l\}& \text{s.t.}\,\,&r_l\geq r_l^{th},\,\,\forall l,
    \end{align}
which is non-convex since $e_l$ is not a concave function of ${\bf \Psi}$. To tackle \eqref{(20)}, we first optimize ${\bf \Psi}$ for $\mathcal{D}_{GPBD}$, which is the most general case, and then discuss how the algorithms can be applied to $\mathcal{D}_{GPD}$ and $\mathcal{D}_{LPD}$. Note that the sets $\mathcal{D}_{GPD}$
and $\mathcal{D}_{GPBD}$ are convex, and solving \eqref{(20)} for a convex set is much simpler than for a non-convex set.

\subsubsection{GP-BD RIS}
To derive a suboptimal solution for \eqref{(20)}, we calculate an appropriate concave lower bound for $e_l$. When $\{{\bf w}\}$ is fixed, the EE of user $l$ is a scaled version of $r_l$. Hence, a concave lower bound for $e_l$ is readily obtained from a concave lower bound for $r_l$. Upon leveraging the bound in \cite[Eq. (71)]{soleymani2023spectral}, we have
\begin{multline}\label{(21)}
r_l\geq \hat{r}_l (\{{\bf w}\})
={r}_l (\{{\bf w}^{(m)}\},{\bf \Psi}^{(m-1)})-b_l
\\
+\frac{2\mathfrak{R}\left\{
\left( {\bf h}_l({\bf \Psi}^{(m-1)} ) {\bf w}_l^{(m)}
\right)^*
{\bf h}_l({\bf \Psi} ){\bf w}_l^{(m)}
\right\}
}{\sigma^2 +\sum_{j\neq l} |{\bf h}_l({\bf \Psi}^{(m-1)} ){\bf w}_j^{(m-1)} |^2}
\\
-a_l\frac{\sigma^2 +\sum_{j\neq l} |{\bf h}_l({\bf \Psi} ){\bf w}_j^{(m)} |^2}{\sigma^2 +\sum_{j\neq l} |{\bf h}_l({\bf \Psi}^{(m-1)} ){\bf w}_j^{(m)} |^2},
\end{multline}
where
\begin{equation}
    b_l=\frac{|{\bf h}_l({\bf \Psi}^{(m-1)} ){\bf w}_l^{(m)} |^2}{\sigma^2 +\sum_{j\neq l} |{\bf h}_l({\bf \Psi}^{(m-1)} ){\bf w}_j^{(m)} |^2}.
\end{equation}
Upon substituting $r_l$ with $\hat{r}_l$, we have
\begin{subequations}\label{(23)}
\begin{align}    
   \max_{ {\bf \Psi} }\, & \min_l \left\{
   \frac{\hat{r}_l}{P_c+\beta{\bf w}_l^{(m)^H}{\bf w}_l^{(m)} }
   \right\}
   \\
   \text{s.t.}\,\,&\hat{r}_l\geq r_l^{th},\,\,\forall l,
   \\
   &\text{Tr}\left(
    \sum_l {\bf F} {\bf w}_l^{(m)}{\bf w}_l^{(m)^H} {\bf F}^H
    ({\bf \Psi}^H{\bf \Psi}-{\bf I}_N)
    \right)
    \leq 0,
   \\
   & {\bf \Psi}={\bf \Psi}^T,
   \label{(23d)}
    \end{align}
\end{subequations}
which is a convex optimization problem and can be solved
efficiently.

\subsubsection{GP-D RIS} 
When $\mathcal{D}= \mathcal{D}_{GPD}$, ${\bf \Psi}$ is a diagonal matrix. In
this case, we replace \eqref{(23d)} by $\psi_{nj}=0$ for all $n\neq j$, which is a convex constraint, and the corresponding optimization problem
is also convex. In general, when $\mathcal{D}$ is a convex set, the overall
algorithm converges to a stationary point of \eqref{(14)} \cite{sun2017majorization}.

\subsubsection{LP-D RIS}
The set $\mathcal{D}_{LPD}$ is non-convex due to the constraint $|\psi_{nn}|=1$, which can be written as the two constraints
$|\psi_{nn}|^2\leq1$ and $|\psi_{nn}|^2\geq1$. The constraint $|\psi_{nn}|^2\leq1$ is convex and can be easily handled. However, the non-convex constraint $|\psi_{nn}|^2\geq1$ makes the corresponding surrogate
optimization problem non-convex. To convexify $|\psi_{nn}|^2\geq1$,
we leverage the approach in \cite{soleymani2023spectral}. Specifically, we approximate $|\psi_{nn}|^2\geq1$ as
\begin{equation}\label{(24)}
  |\psi_{nn}^{(m-1)}|^2-2
  \mathfrak{R}\left\{\psi_{nn}^{(m-1)^*}(\psi_{nn}-\psi_{nn}^{(m-1)})
  \right\}
  \geq1-\epsilon,  
\end{equation}
where $\epsilon > 0$. Upon employing \eqref{(24)}, we obtain the convex surrogate problem
\begin{subequations}\label{(25)}
\begin{align}    
   \max_{ {\bf \Psi} }\, & \min_l \left\{
   \frac{\hat{r}_l}{P_c+\beta{\bf w}_l^{(m)^H}{\bf w}_l^{(m)} }
   \right\}
   \\
   \text{s.t.}\,\,&\hat{r}_l\geq r_l^{th},\,\,\forall l,
   \\
   &\eqref{(24)},|\psi_{nn}|^2\leq1,\,\,\forall n,
   \\
   & \psi_{nj}=0, \,\,\forall n\neq j,
   \label{(25d)}
    \end{align}
\end{subequations}
which can be easily solved. The solution of \eqref{(25)}, ${\bf \Psi}^{(\star)}$, may be non-feasible because of the relaxation in \eqref{(24)}. To tackle this issue, we first normalize ${\bf \Psi}^{(\star)}$ as $\psi_{nn}=\frac{\psi_{nn}^{(\star)}}{|\psi_{nn}^{(\star)}|}$. Then, we update ${\bf \Psi}$ according to \cite[Eq. (50)]{soleymani2023spectral} to obtain a non-decreasing sequence for the max-min EE, which guarantees
the convergence of the proposed algorithm.

\section{Numerical results}

We evaluate the proposed algorithm through Monte Carlo simulations. The RIS-related links are assumed to be in line of sight (LoS) and to follow a Rician distribution. Thus, the channels ${\bf F}$ and ${\bf f}_l$ can be modeled according to \cite[Eqs. (55)-(57)]{pan2020multicell}. We assume that the Rician factor for the channels ${\bf F}$ and ${\bf f}_l$ is $3$, similar to \cite{pan2020multicell}. Furthermore, the direct link between
the BS and each user is assumed to be in non-LoS (NLoS) and to follow the Rayleigh distribution. The large-scale path loss of each link is calculated based on \cite[Eq. (59)]{soleymani2022improper}. The static powers of D-RIS and BD-RIS are calculated based on \eqref{(12)} and \eqref{(13)}, respectively. We assume $P^D_{RIS,0} = P^{BD}_{RIS,0}= P_{RIS,0}=0.01$ W and $P^D_{RIS,n} = P^{BD}_{RIS,n}= P_{RIS,n}$. Additionally, the other simulation parameters are chosen as in \cite[Table II]{soleymani2022improper}.

\begin{figure}[t]
    \centering
           \includegraphics[width=.45\textwidth]{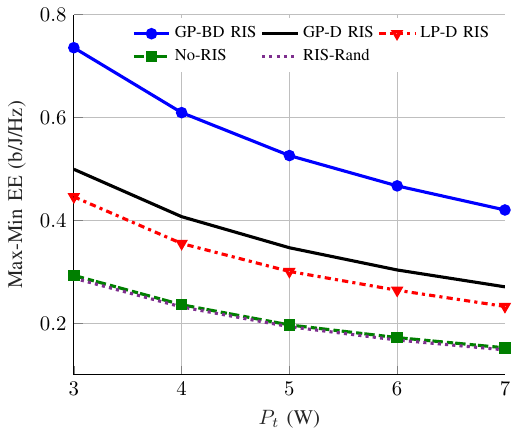}
    \caption{Average max-min EE versus 
 $P_t$ for $P_{RIS,n} = 1$ dBm, $K=5$, $L=5$, $N=20$ and $P=10$ dB.}
	\label{Fig1} 
\end{figure}
Fig. \ref{Fig1} shows the average max-min EE versus $P_t$ for $P_{RIS,n} = 1$ dBm, $K=5$, $L=5$, $N=20$ and $P=10$ dB. This figure illustrates that all the RIS architectures considered in this paper substantially enhance the max-min EE if the RIS elements are optimized based on the proposed schemes. Indeed, even though the constant power consumption of the system without RIS is assumed to be lower than the constant power of the RIS-aided systems, all the RIS architectures significantly improve the EE of the system. Moreover,  GP-D RIS outperforms LP-D RIS, while GP-BD RIS significantly outperforms all the other schemes.

\begin{figure}[t]
    \centering
           \includegraphics[width=.45\textwidth]{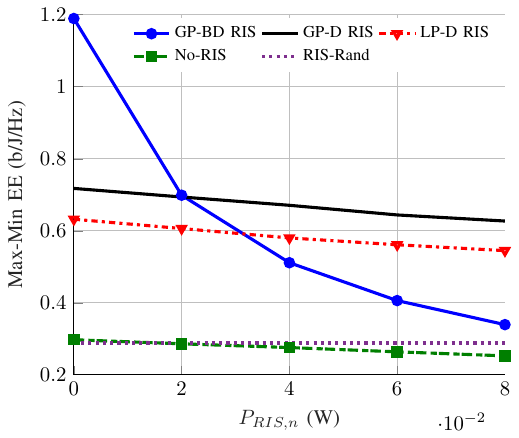}
    \caption{Average max-min EE versus 
 $P_{RIS,n}$ for $P_t=3$ W, $K=5$, $L=5$, $N=40$ and $P=10$ dB.}
	\label{Fig2} 
\end{figure}
Fig. \ref{Fig2} shows the average max-min EE versus $P_{RIS,n}$ for $P_t=3$ W, $K=5$, $L=5$, $N=40$ and $P=10$ dB. As the circuit power of each RIS/circuit element increases, the EE of different RIS architectures decreases. The EE of GP-BD RIS is more affected by increasing $P_{RIS,n}$ since this architecture has many more circuit elements than LP-D RIS and GP-D RIS. Interestingly, GP-BD RIS provides a lower max-min EE than the D-RIS architectures when $P_{RIS,n}$ is not sufficiently low. Moreover, GP-BD RIS may not improve the EE when $P_{RIS,n}$ is too high. This result shows that the EE of GP-BD RIS drastically depends on the energy efficiency of the hardware implementation and circuits design. Indeed, the static power consumption of the RIS/circuit elements has a more significant impact on BD-RIS than D-RIS because BD-RIS has a large number of circuit elements.

\section{Conclusion}
In this paper, we proposed energy-efficient schemes for three RIS architectures, i.e., LP-D, GP-D, and GP-BD, for application to MU-MISO BCs. We showed that these three architectures can significantly increase the max-min EE if the RIS elements are optimized according to the proposed algorithms and if the static power consumption of the RIS elements is not high. Moreover, GP-BD RIS substantially outperforms GP-D RIS and LP-D RIS if the circuit elements of GP-BD RIS are very energy efficient and consume a low static power. Otherwise, GP-BD RIS may perform worse than its diagonal counterparts. Furthermore, GP-D RIS is slightly more energy-efficient than LP-D RIS. Additionally, the EE benefits of RISs may be negligible if we do not appropriately optimize the RIS elements.
\section*{ACKNOWLEDGMENT}
The work of I. Santamaria was funded by MCIN/ AEI/10.13039/501100011033, under Grant PID2022-137099NBC43 (MADDIE), and by European Union’s (EU’s) Horizon Europe project 6G-SENSES under Grant 101139282. The work of E. Jorswieck was supported in part by the Federal Ministry of Education and Research (BMBF, Germany) through the Program of Souverän. Digital. Vernetzt. joint Project 6G-RIC, under Grant 16KISK031, and by European Union’s (EU’s) Horizon Europe project 6G-SENSES under Grant 101139282. The work of M. Di Renzo was supported in part by the European Commission through the Horizon Europe project titled COVER under grant agreement number 101086228, the Horizon Europe project titled UNITE under grant agreement number 101129618, and the Horizon Europe project titled INSTINCT under grant agreement number 101139161, as well as by the Agence Nationale de la Recherche (ANR) through the France 2030 project titled ANR-PEPR Networks of the Future under grant agreement NF-YACARI 22-PEFT-0005, and by the CHIST-ERA project titled PASSIONATE under grant agreements CHIST-ERA-22-WAI-04 and ANR-23-CHR4-0003-01.

\bibliographystyle{IEEEtran}
\bibliography{ref2}

\end{document}